\documentstyle[aas2pp4,flushrt]{article}

\slugcomment{Submitted to {\it The Astrophysical Journal} }
\lefthead{LINN, ET AL.} 
\righthead{CLUSTERS OF GALAXIES IN N-BODY EXPERIMENTS}
\begin{document}

\title{The  Ellipticity and  Orientation of  Clusters  of  Galaxies in
N-Body Experiments}

\author{Randall       J.         Splinter\altaffilmark{1,2},    Adrian
L. Melott\altaffilmark{3}  Angela  M.   Linn\altaffilmark{3},  Charles
Buck\altaffilmark{3}, and Jeremy Tinker\altaffilmark{3} }

\affil{E-MAIL: randal@ccs.uky.edu, melott@kusmos.phsx.ukans.edu, \\
linn@kusmos.phsx.ukans.edu, buck@kusmos.phsx.ukans.edu, \\
tinker@kuplas.phsx.ukans.edu}

\begin{abstract}

In this study  we  use simulations  of 128$^3$ particles  to study the
ellipticity  and orientation    of  clusters  of  galaxies in   N-body
simulations of differing  power-law initial spectra  ($P(k)\propto k^n
,n = +1,  0,  -1, -2$), and density   parameters ($\Omega_0 = 0.2$  to
1.0),  in a controlled way, based  on nearly  3000 simulated clusters.
Furthermore, unlike most theoretical  studies we mimic  most observers
by removing all particles which lie  at distances greater than $2 {\rm
h^{-1}}$ Mpc from the cluster center of mass.

We  computed the axial ratio and  the principal axes using the inertia
tensor of  each cluster.  The  mean ellipticity of  clusters increases
strongly  with increasing  $n$.  We  also find  that clusters tend  to
become more spherical at smaller radii.

We  compared the  orientation  of  a  cluster  to the   orientation of
neighboring  clusters  as a function  of   distance (correlation).  In
addition, we considered  whether a cluster's  major axis  tends to lie
along the line  connecting it to a neighboring  cluster, as a function
of   distance (alignment).   Both   alignments and   correlations were
computed in three dimensions and  in projection to mimic observational
surveys.  Our results show that  significant alignments exist for  all
spectra at small separations ($D < 15 {\rm h^{-1}}$ Mpc) but drops off
at larger  distance  in a strongly   $n-$dependent way.  Therefore the
most  useful  study for observers  is the  variation of alignment with
distance. Correlations exist but are a weaker effect.
	
We found that differences in $\Omega$ had no measurable effect on mean
ellipticity,  and  a      weak effect  on    cluster   alignments  and
correlations.  Biasing was able to  totally hide the effect of greater
nonlinearity. Therefore, we suggest that  any effort to probe $\Omega$
in this  manner be abandoned unless  it can be unambiguously proven to
exist on smaller scales.  However, there are systematic effects due to
the primordial spectral index, $n$.   Our results suggest that cluster
ellipticity and the scale dependence  of cluster alignments probe  the
primordial   power spectrum independently  of  the  parameters  of the
background  cosmology.   Future   work  should  concentrate  on  these
parameters.

\altaffiltext{1}{Department of Physics and Astronomy,
University of Kentucky, Lexington, KY 40506-0055.} 
\altaffiltext{2}{Center for Computational Sciences,
University of Kentucky, Lexington, KY 40506-0045.}
\altaffiltext{3}{Department of Physics and Astronomy,
University of Kansas, Lawrence, KS 66045.} 

\end{abstract}

\section{Introduction}

A common theoretical picture  of large-scale structure formation holds
that  a hierarchical  (bottom-up)  scenario   will result  in   galaxy
clusters of  lower   ellipticity  and possessing little  evidence   of
preferred orientation with    neighbors.  Alternatively,  if   pancake
structures  exist as predicted  by Zel'dovich (1970),  then a top-down
emergence  of structure should result    in elliptical clusters   with
statistically    significant   alignment correlations  with neighbors.
Recently,  these  theoretical models  have been  effectively combined,
implying a transition based on spectrum should be present (e.g. Melott
and Shandarin, 1993; Pauls and Melott 1995; Bond et  al. 1996).  It is
supposed that a model  which  is characterized  by high power  in  the
short-wavelength end  of  the spectrum should  be  expected to exhibit
more hierarchical traits compared to a long-wavelength dominated model
and   its more top-down mediated   structure.   Furthermore, in a  low
density  universe  structure formation  slows near a  redshift $z \sim
1/\Omega$.   Thus the cluster may undergo  a long period of relaxation
without additional infall after its  initial formation, and may become
more spherical.   For these  reasons,  many have hoped  that clusters,
although nonlinear objects, may  probe initial conditions. They are at
the  borderline of  scales  where  hydrodynamics  are  thought  to  be
essential (Summers et al. 1995).

Carter  \& Metcalfe (1980)  showed  that the distribution of  galaxies
within a  cluster is   usually  not spherical,but    highly elongated.
Additionally, a number of studies (Rood, et  al 1972, Gregory \& Tifft
1976, and  Dressler 1981) showed  that  the elongation was  not due to
rotation.   Binggeli (1982), through  his study of Abell clusters form
the Palomar Sky Survey, found evidence that galaxies separated by less
than $\sim 30$ Mpc  exhibit a strong tendency  to point at one another
(we  call  this ``alignment'' hereafter).   He    also found that  the
orientation of a cluster  was dependent upon  the distribution  of the
surrounding clusters, thus arguing  for an anisotropic distribution of
clusters  on  a scale  of $\sim  100$ Mpc.  However,   in a repeat and
extension  of Binggeli's methods,  Struble   and Peebles (1985)  found
little   evidence  for large-scale      alignment   of clusters     of
galaxies.  West  (1989b) on  the   other hand confirmed  the  Binggeli
result.

In more recent work, de Theije, et al  (1995) (hereafter TKK) in their
study of 99 low-  redshift Abell clusters  found that the distribution
of cluster ellipticities  had a peak near  $e \sim 0.4$ with a maximum
of  $e \sim 0.8$. Rhee,  et al  (1992)  found evidence for large-scale
alignment in a study of 107 rich clusters from the Palomar Sky survey.
West  and colleagues have  also found significant  evidence to support
this viewpoint.  West,  et al  (1995) found a  marked anisotropy based
upon their  studies  of  the  X-ray  distribution of  substructure  in
clusters of  galaxies from the  Einstein  observatory.  In  1989, West
generated a catalog  of 48 superclusters based  upon a large number of
Abell clusters to study the shapes and orientations of those clusters.
He  found  that  the  axial  ratios tended to   fall  around  3:1:1 or
4:2:1. Furthermore, he  presented evidence that on  scales of as large
as $\sim 60 {\rm h^{-1}}$ Mpc  there is a strong  tendency for them to
be aligned. There has thus developed over the last few years a picture
in which clusters  are  clearly elliptical and exhibit   statistically
significant alignments    toward   other  members  of   the     parent
supercluster. Recently Plionis (1994)  has completed the largest up to
date study of cluster alignments. He finds nearest neighbor alignments
up to scales of  15 $h^{-1}$ Mpc at the  2-3 sigma significance level,
while weaker alignments are detected on scales up  to 60 $h^{-1}$ Mpc.
Since his  cluster sample is not  volume limited nor redshift complete
the alignments he detects are likely to reflect a real effect.

Using N-body studies of a CDM universe,  West, et al (1991) (hereafter
WVD) found that   on scales of   order  $10-15 {\rm  h^{-1}}$ Mpc  the
principal axes of neighboring  clusters were clearly aligned. When the
sample  of  clusters  is    limited   to those clusters    ``within  a
supercluster'', the alignments extend  to distances as large as  $\sim
30  {\rm  h^{-1}}$ Mpc.  TKK   found  evidence, based  upon  numerical
simulations of $\Omega=0.2$  CDM  models, that clusters  tended  to be
more spherical in   low-omega models. This  confirmed the  theoretical
work of van Haarlem  and van de Weygaert  (1993) who came to the  same
conclusion  based  upon  N-body simulations  of   a CDM universe.  Van
Haarlem and Van  de Weygaert also  found that their simulation results
are  consistent with  a picture   in which alignments  arise from  the
infall of substructure   along a filament,  anticipating  a finding of
West, et    al  (1995). Earlier,    Shandarin and  Klypin  (1984)  had
explicitly interpreted the  results of  their  simulations as  showing
that clusters were the result of flows along filamentary structures.

We are  trying here to  systematize and probe  the significance of the
N-body studies.  In some  studies,  cluster separations were so  large
relative to the simulation volume that boundary conditions are clearly
a problem.  In  other cases, the  quantities studied,such as alignment
in three dimensions are clearly not measurable in  real data given the
precision of  non--redshift  distance measures and distortions  due to
peculiar  velocity  (e.g.   Praton and  Schneider  1994).  In studying
popular scenarios, there is a tendency to  vary multiple parameters at
once.  For example,   TKK and van  Haarlem  and van de  Weyaert (1993)
simultaneously varied $\Omega_0$,  the  bias factor, and the   initial
power spectrum, since the  low $\Omega$ CDM  universe has a  different
transfer function and presumed bias.

Given the variety of  possible theories, it is  important to know what
parameters are  being probed and what are  not, rather than relying on
case--by--case checks of  agreement.  Ultimately, it will be necessary
to do simulations large enough to  sample long--wave power but able to
resolve  hydrodynamic scales.  At  that time  there  can be a physical
basis  for   assuming  some  bias between   mass  and  galaxies. Since
dissipation erases information,  our study will  set an upper limit on
the kind  of information   that can  be recovered  in   global cluster
studies. This study will clarify what can and cannot be probed by such
cluster statistics.

\section{N-Body Simulations}

The N-body simulations used  were generated using a Particle-Mesh (PM)
code  using a staggered--mesh scheme   Melott (1986).  Our simulations
consisted of $128^3$ particles on a comoving $128^3$ grid (for details
see    Melott and Shandarin 1993).  To    perform the data analysis we
considered only a $64^3$ subset of the  original $128^3$ particles but
we have  checked a sample of our  results for  consistency with a full
count.   Even  for  our smallest   clusters   there was no significant
difference in our main  diagnostics between the  subset and the use of
all particles from the simulation.   We ran simulations for four power
law spectra,   $n = 1, 0, -1,   -2$,  in both  high-  and low-$\Omega$
universes.  Four  realizations  of  each  of the  eight above  initial
conditions  were performed.  These realizations  were  then studied at
two   different  evolutionary     timesteps, $k_{n\ell}=8 k_f$     and
$k_{n\ell}=4 k_f$,  where    the non-linear wavenumber;   $k_{nl}$  is
defined by

\begin{equation}
\sigma^2 = a^2 \int^{k_{nl}}_0 P(k) d^3k = 1,
\end{equation}

\noindent where $P(k)$  is the initial  power spectrum of  the density
fluctuations, and $a$ is the cosmic expansion factor.

In the  high-$\Omega$ case,  $\Omega_0  = 1$ for   all cases.  For the
low-$\Omega$  simulations, $\Omega_0 =    0.2$ for all power  laws  at
$k_{n\ell}=4 k_f$ at the earlier stage $k_{n\ell}=8 k_f$, $\Omega$ had
the values $0.707, 0.609, 0.487$ and  $0.347$ for $n =  1, 0, -1,$ and
$-2$, respectively, and $k_f$ is the wavenumber of the fundamental.

Power--law spectra  are not  realizations of any  particular scenario,
but are extremely useful as probes of the physics of clustering. Since
the initial spectrum is featureless, we can scale from the correlation
length of mass as set to $\sim 5h^{-1}$ Mpc, setting the free ``bias''
parameter $b=1$. Our goal here is to find out what statistics {\it can
be} useful discriminators of cosmological models.

From   each    realization,   clusters   were   identified   using   a
``friends-of-friends'' technique.  Any  points closer than the linking
distance to  a given point were considered  linked.  The linked points
were then in turn checked for points linked  to them, thereby creating
clusters. A linking distance of $l=0.5$ grid  cells was chosen because
it leads   to clusters of approximately    the same mass   as an Abell
cluster  $\sim 10^{15}  {\rm  M_\odot}$ assuming  $\Omega=1$.  Table 1
contains the resulting  box   sizes  for  each  of the   models  being
considered  here.   For  stage  $k_{n\ell}=4  k_f$,   the ten  largest
clusters were selected for examination,  ten being about the number of
clusters expected in that volume of space  (Batuski et al.  1987).  In
addition Hoessel et  al (1980)  found  the observed number density  of
clusters  of richness  class $R \leq   1$ to be  $n  \approx  6 \times
10^{-6} h^3 {\rm Mpc^{-3} }$.  Bahcall (1988) argued a more reasonable
value when galactic obscuration is taken account is $n \approx 10^{-5}
h^3 {\rm  Mpc^{-3} }$.   Given  the sizes of  the computational volume
given in Table 1, which are typically $\sim 100 h^{-1}$ Mpc on a side,
our use of  10  clusters per simulation   is justified.   Since  stage
$k_{n\ell}=8 k_f$ represents  a  volume of space $2^3$  times  greater
than $k_{n\ell}=4 k_f$, we   chose eight times  as many  clusters (the
eighty largest clusters) from each realization in  that stage.  In all
then, our  study contains 2880 clusters,  many  more than any previous
study.  None have less than 40 simulated particles, and most have many
more.

We wish to stress that in this study  we are restricting our attention
to    a  controlled  and    systematic  study  of  the   properties of
collisionless systems, i.e. dark matter, presuming that it consists of
a population  of  weakly  interacting particles.   It  has  been shown
(Suisalu and Saar  1996) that unless  the comoving softening length is
approximately equal   to the   mean interparticle distance,   spurious
gravitational  collisions occur  which  scatter the particles  off one
another  (as opposed to the mean  field). In this   limit a PM code is
much faster than anything else.

We will  thus  be unable  to probe  the inner  parts  of clusters; our
resolution   is   limited    to  an   equivalent   0.5  Mpc    in this
series. However,  in order to  model these inner regions correctly, we
would need to include hydrodynamics (e.g.  Evrard et al. 1993, Summers
et  al 1995).  By forgoing  this  for now,  we  are able  to conduct a
large, controlled, systematic study of the  overall shapes of clusters
free of two--body relaxation, as implicit in PM codes (e.g. Peebles et
al. 1989).

\section{Average Ellipticity}

\subsection{Method of Computing Ellipticity}

The clusters as we find them  may have arbitrary shapes, including the
filaments that connect   them  and outliers  just in   the process  of
merging.  For  this reason, we   follow most observers'  practice, and
inscribe a sphere of radius 2.0$h^{-1}$ Mpc in radius, centered on the
center of mass  of the cluster. This  choice is  motivated by distance
scales  which can be  separated observationally,  (cf.  Bahcall \& Cen
1993, Rhee, et   al 1991a, Rhee, et   al 1991b, Rhee  \& Latour  1991,
Binggeli 1982 ) but  is nonetheless somewhat arbitrary.   In computing
ellipticity and orientation we use  only this portion of the  cluster,
as in previous observational studies of this type.

The  ellipticity of a  cluster was found  using second  moments of the
cluster mass.  This  method  is similar  to  that outlined in  TKK and
Plionis, Barrow \& Frenk (1991).  With respect to  the center of mass,
the tensor for the two dimensional case is defined as

$$I_{ij}=\Sigma \chi_i \chi_j m \hskip .5in (i,j,=1,2)\eqno (1)$$

\noindent where the sum is over all points, $\chi_1=x$ and $\chi_2=y\;
.$ This  tensor  has the same   eigenvectors as  the  standard inertia
tensor but its eigenvalues are  related to axis  length. For a uniform
ellipsoid the ratio of eigenvalues is the square of the ratio of axes.
In our simulations, the points were all of equal mass,  so $m$ was set
to  1.  This can  be  easily extended to   three dimensions by letting
($i,j=1,2,3$), where $\chi_3=z .$  $I_{ij}$ was diagonalized to find
the   eigenvalues   and eigenvectors.   The  eigenvalues $\Lambda_1,\;
\Lambda_2,\;\Lambda_3$          were       sorted        so       that
$\Lambda_1\geq\Lambda_2\geq\Lambda_3$               (or           just
$\Lambda_1\geq\Lambda_2,$    for  two   dimensions).   For   the   two
dimensional case, the axial ratio was defined as

$$e=\sqrt{\Lambda_2/\Lambda_1}\eqno (2)$$

\noindent For the three dimensional case, two axial ratios were found:

$$e_2=\sqrt{\Lambda_2/\Lambda_1}\; \; ,\; \;
e_3=\sqrt{\Lambda_3/\Lambda_1}\eqno (3)$$

In order to compare our results with those of TKK, the ellipticity was
defined for the two dimensional case as

$$\epsilon=1-\Lambda_2/\Lambda_1\equiv 1-e^2\eqno (4)$$

The  results in the 2-D  case  are those for  projections of clusters.
Because   the individual   clusters  are oriented   (as we  confirmed)
randomly with respect to the $x$, $y$,  and $z$ axes, just as physical
clusters are oriented  randomly with respect  to the plane of the sky,
this assures an unbiased data set. For the  two dimensional case, each
cluster was  projected onto the $xy$ and  $yz$, planes, since only two
projections are independent in $e$.

\subsection{Average Ellipticity and Initial Conditions}

In Figure 1a and 1b we plot $e_2$ and $e_3$.  In Figure 1c we plot $e$
for the 2D projections.  The error bars are  the standard deviation of
the  {\it mean} value for each  of the  realizations. The single error
bar in the lower left  corner represents the dispersion in ellipticity
for a typical cluster.  We are probing whether  systematic differences
in the average  values    can give cosmological   information,   given
datasets the  size of our simulation  volumes.  Further fine tuning of
observables  will require imitating particular survey characteristics,
which is outside  the purpose of  this paper.  However, we can already
make a number of unambiguous statements:

(a) Clusters are not spherical; in general they are triaxial.

(b)  We  can  check  for  consistency by    examining results  in  the
scale--free $\Omega_0=1$ case. There are low--significance differences
between  the stages for $n=1$ in  the axial ratio, smaller for smaller
$n$. We have confirmed that  these reflect  resolution effects in  our
code, which affect the earlier stage $k_{n\ell}=8 k_f$. The difference
is a measure of how much numerical effects  limit the precision of our
answers. The change  in error bars  is due to a  change  in the sample
size. At this point one might be  worried that the $n$-dependent trend
might  be  simply a numerical  artifact.  We  have tested  for this by
computing the ellipticities  at   a later  stage ($k_{nl}=4 k_f$)   of
evolution where resolution is  not  problem.  These results are   also
shown in figures 1a, 1b, \& 1c. At this later  stage the trends remain
thus they cannot be due to limited dynamical range in the simulation.

(c) There is no significant  difference in ellipticity between low and
high $\Omega$.  This appears   to contradict  the  results of  TKK and
others, but we believe we understand the reasons, as explained below.

(d) There is a systematic trend with spectral index: large $n$ implies
more anisotropic  clusters. Thus,  although  it is  hopeless  to probe
$\Omega_0$ with cluster ellipticities, it may  possibly probe $n$ more
clearly.

(e) The  dispersion  in  ellipticity of clusters    is large within  a
cosmological  model,  and large  within a  sample.   At  least 100
clusters will  be needed to   establish   the result with   sufficient
precision.

(f) Clusters are more isotropic at smaller radii. (Figure 2)

Now,  we  compare those conclusions  with other  studies.  (a) and (f)
agree with all  others where they were  checked. (b)  is a consistency
test not performed elsewhere.

Conclusion (c) appears to conflict with TKK, as  well as Evrard et al.
(1993).  These  authors studied Cold Dark  Matter models and concluded
their    low  $\Omega_0$ CDM   models   produced  much less  elongated
clusters. There are numerous differences with our work, including many
fewer   clusters,   smaller   total     volumes studied,   constrained
realizations of random fields,  varying size of clusters, not choosing
to   excise  the central  region of   the cluster, etc.   However, one
consistent difference  is  the choice  of {\it  biasing}. In all these
studies  a     bias factor $b$   (usually  1.7-2)   is used   with the
$\Omega_0=1$ models. (A  typographical error in  TKK  was confirmed in
conversation   with  Katgert.)  $b$   is    typically defined as   the
$1/\sigma_8$, where $\sigma_8$ is the RMS  mass density fluctuation in
a 8  Mpc sphere.  In all  cases,  these CDM  studies have  a different
ratio of cluster mass to  mass scale of nonlinearity  for low and high
$\Omega_0$.   Further support for this point  of view can  be found by
comparison with West et al.  (1989) (hereafter  WDO) who (similarly to
us) found no  difference between their  $n=0$ low $\Omega_0$ and  high
$\Omega_0$ ({\it unbiased}) models.

Also, when TKK  went from high  to low $\Omega_0$   they used the  CDM
transfer  function for   each  case.  Low $\Omega_0$  CDM   has a more
steeply negative initial power  spectrum at  cluster scales than  high
$\Omega_0$.  Our observed trend (d) would explain  the TKK result: the
change  in  ellipticity  is a   result of  a  different  initial power
spectrum, not the value  of $\Omega$. Their result  is true  for these
particular CDM alternatives, but does not generalize.

Evrard et al. (1993) examined contours at a much smaller radii than we
did, so  the results cannot be  directly compared. The difference they
found   (more spherical inner contours in   low $\Omega$) could be due
either to the improved modeling  due to including hydrodynamics, or to
spurious scattering and two--body relaxation (as Suisalu and Saar 1996
found in  $P^3M$  codes;   see also   Merritt 1996),  or  is  simply a
phenomenon of small radii only.

Conclusion (d) does not appear in the conclusions of West et al, 1989,
hereafter WDO and Efstathiou et  al.  (1988), who found no significant
trend in ellipticity with power--law index.  However,  WDO had too few
clusters (20 per spectral type) to see  these trends.  Our results are
consistent  with theirs.  In Figure 1  the standard  deviation for one
cluster,  shown as  the  single error  bar in  the lower left  corner,
(rather than the mean in  the volume) is  about an order of  magnitude
larger.  It is clear  that large ensembles  such as those to come from
the Sloan  Sky Survey  are needed to  measure these  effects.   On the
other  hand, Efstathiou et  al.  (1988) who had  enough objects to see
this made  a fatal cut:  they excluded  from consideration as clusters
any $n$--body  clumps which were not ``smooth'',  i.e.  had no central
mass    condensations   as measured  by  asymmetry     at a variety of
radii. Since  the asymmetries  are    typically generated by   mergers
falling  in along bridges that connect  clusters (Shandarin and Klypin
1984, TKK) cutting out these ``unsmooth'' clumps means eliminating the
main  source  of   anisotropy.   Most real  clusters  are  not smooth.
X-imaging results from the {\it Einstein} satellite first revealed the
complexity of the  intracluster medium (Forman,  et al 1981;  Henry et
al. 1981). Since then both optical (Geller  \& Beers 1982; Baier 1983;
Beers, et al.   1991; Bird 1993, 1994a,b)  and X-ray (Jones  \& Forman
1992;  Davis  \& Mushotzky   1993; Mohr, et    al.  1993) studies have
continued to support  the  presence  of significant  substructure   in
galaxy clusters (for a  theoretical perspective see Dutta 1995; Crone,
et al. 1995;  Splinter \& Melott 1996).  These results contradict many
early studies  in which smooth  configurations were assumed (cf.  Kent
\& Gunn 1982;  Kent \& Sargent  1983).  Thus, Efstathiou  et al (1988)
eliminated the most interesting cases.

The trend we  observe, that of  increasing ellipticity with increasing
$n$, can  be understood by  considering the  mass function  (Press and
Schechter 1974).  For  larger $n$, the mass  function  dives down more
steeply at small masses.  Thus, for large $n$,  mergers are more often
between  large  clumps, which  significantly perturbs  the  shape. For
smaller $n$, the clusters grow by accretion of smaller clumps, tending
toward  accretion of homogeneous mass  as $n$ approaches -3. Accretion
of many small clumps tends to  leave the cluster shape undisturbed, as
seen in our results.

We confirm (Figures  2a,2b \& 2c) that  clusters become more isotropic
in   their shape  at smaller  radii.   The figure   also compares (for
spectral index  $n=-1$)   an $\Omega=1$, $b=1$  simulation  and  a low
$\Omega$,  $b=1$ simulation.   This bias is   defined operationally by
comparing the 10 largest clusters in a  $k_{n\ell}=4 k_f$ box with the
10 largest clusters in  a $k_{n\ell}=8 k_f$ box  (which would be taken
to contain 80 clusters if $b=1$).  We  find essentially identical runs
of ellipticity with radius, suggesting they cannot be distinguished by
this  procedure.  Biasing appears able  to  imitate the effect of more
nonlinearity for cluster ellipticity.

\section{Alignments \& Correlations}
\subsection{Alignment}

We use alignment  of a cluster to refer  to whether the principal axis
points along the   line joining it   to a neighboring   cluster.  This
continues investigation into suggestions  of Tifft (1980) and Binggeli
(1982). The principal axis is  determined from the eigenvectors of the
inertia tensor, found    above.  The eigenvector  associated with  the
largest eigenvalue is the longest principal axis of the cluster.

The cosine of the angle $\alpha$ between a cluster's longest principal
axis and the  line connecting it to a  neighboring cluster was plotted
against the  distance between the  two  clusters.  In order to  remove
effects of  boundary  conditions,  the cluster--cluster   distance was
never larger than $L/4$.

We attempted  to quantify  this     observation in three  ways.     We
determined the mean and standard deviation of $cos\; \alpha$ using all
the  data points.  $<cos  \alpha>$ is  0.5 for  a random uniform three
dimensional  distribution, and   $<\alpha>$  is $45^\odot$ for  a  
random two
dimensional  distribution.  We  also  examined projected angles, since
the full angle is not observable.

\subsection{Correlation}

We use  here ``correlation'' of two clusters  as  a measure of whether
their longest principal axes point in the  same direction.  This would
be likely if ellipticities were  induced by tidal forces, rather  than
merger history as in alignment.   This quantity was calculated in much
the same way as in the alignment.  The  cosine of the angle $\beta$ by
which  the principal axes  of two  clusters  are separated is  plotted
versus   distance in  Figures  4.  The  values  of  $<cos \beta>$  and
$<\beta>$ are 0.5 for a three and two dimensional random distribution,
respectively.

\subsection{Statistical Analysis of the Alignment and Correlation Results}

Figures 3  and 4 give  a visual impression of  the significance of the
alignments and correlations  in our  datasets, but visual  impressions
can hardly be considered a  robust method of  estimating the degree to
which  alignments or  correlations exist  in the datasets.   To make a
more definitive  statement  concerning  the statistical likelihood  of
alignments or  correlations existing in   the datasets we  introduce a
well defined  null hypothesis.  The null  hypothesis we choose is that
the given datasets   are consistent with  uniformly distributed random
alignments  or  correlations.  To test this  null  hypothesis we use a
Kolmorogov-Smirnov (hereafter  KS) test (e.g. Press,  et al 1987). Our
choice   of the  KS   test is  based upon   our  desire to  avoid  any
discreteness  effects which can  be introduced  by  use of binning the
data for use in the $\chi^2$ test.

We  first  bin  the data into   distance   bins and  then  analyze the
distribution of  clusters at each of  the distance bins, since we want
to test whether  the distribution of clusters  at all  distance scales
exhibits any  alignments or correlations.    For each of the  distance
bins  a KS test  is performed.  For small  distances there are often a
small   number   of clusters.  For    this  reason  we  exclude  stage
$k_{n\ell}=4 k_f$ from this analysis.  The  number of clusters in each
distance bin at    this stage   was   often $\le  10$  making   strong
statistical statements difficult. We  find significant  alignments and
possibly one overall pattern that will help  to discern the background
cosmology.   The results  for three-dimensional alignments  as well as
the more observationally relevant two-dimensional projected alignments
are  shown in  tables  2,  3,  4, and  5.  The  quoted errors  are the
dispersion in the  mean for the  sample  size shown, 320 clusters  per
spectral index and background density.

\subsection{Alignments, Correlations and the Initial Conditions}

Despite the  early   contradictory   observational evidence  for   the
large-scale alignments of galaxy  clusters (cf. Binggeli 1982, Struble
\& Peebles 1985, and West 1989b) it is now clear that there does exist
evidence for the  large-scale alignment of  clusters (cf. Rhee, et al
1992 and West, et al 1995). In addition, West, et al (1991) (hereafter
WVD) found evidence  based upon N-body studies  in a CDM universe that
there exist alignments on scales up to 15 h$^{-1}$ Mpc, and when those
clusters   are limited to  members  of  a  supercluster the alignments
stretch up to  nearly 30 h$^{-1}$ Mpc.  This  is in agreement with
the results we find here.

It appears from  our   data that it   may be  difficult to  probe  the
background cosmology  using alignments and correlations. There appears
to be  a very weak trend that  as $\Omega$ is lowered  more alignments
and correlations are seen, but we stress it is very weak. 

On  the other   hand, there is  hope  for   using the alignments   and
correlations to probe the  primordial  spectrum.  The trends here  are
much more straightforward;  as  $n$ is   made more  negative there  is
increasing alignments and correlations between  clusters. In all cases
there is  alignment  for $D<15h^{-1}$Mpc. Table   2 (and Table  4, for
projected  angles  against real    distances)  show alignment  out  to
$30h^{-1}$Mpc for $n=0$, and  on all scales  for $n\leq -1$. Alignment
is slightly stronger in low $\Omega$  models, but the difference is so
small that we do not consider it promising. 

Cluster angle   correlations display more noise  with  a strong signal
only for small separations ($D  < 15  {\rm  h^{-1}}$ Mpc), while  both
$n=-2$ \& $-1$ display moderate signals on larger scales. The strength
of the signals does not appear strong enough to be useful.

Cluster alignments appear to probe the primordial power spectral index
nearly independent of $\Omega$,  and we recommend probing  its angular
separation dependence in two dimensions along with a three dimensional
study.

\section{Discussion and Conclusions}
	
We have looked at the average  ellipticity of clusters and projections
of     clusters,  cluster-cluster  alignment,   and    cluster-cluster
correlation in an  attempt to distinguish between  N-body cosmological
models of differing $\Omega$ and initial power-law spectra.

The  ellipticity of  clusters and projections   of  clusters shows  no
significant relation to $\Omega_0$ and does not  appear to change with
time.  Likewise, Walter \&  Klypin (1995) conclude that elongation  of
clusters has not changed with time.  We do find that clusters are more
elliptical as  we go to   larger $n$ (that   is, more  power on  small
scales). We attribute this to merger of larger fragments.

We confirm a trend in axial ratio with $n$;  since this is independent
of $\Omega_0$, it has some hope of being  a robust measure of spectral
index.  We conclude that  the spectral index  $n$ is probed by cluster
ellipticity,  but  not $\Omega_0$.    The  probing of    $\Omega_0$ by
dissipative processes  may be possible,  in some  convolution with the
timescale of  the  background  cosmology.   However,  in  general  our
results  set somewhat pessimistic upper limits  on how much background
information on    cosmology  can come  from    cluster studies,  since
dissipation generally destroys information.

Cluster-cluster alignment, a measure of whether  a cluster is pointing
to a neighboring cluster, shows only  a strong relation to the initial
power-law spectra, with clusters tending to be more aligned when there
is  more large-scale power (i.e.    with decreasing $n$).  There is  a
relation between   the initial  power-law spectra   and the amount  of
alignment in our N-body simulations.  Alignment exists for close pairs
for all spectral  indices, but extends to  larger and larger scales as
$n$ decreases. Van Haarlem \& Van  de Weygaert (1993) have argued that
the orientation of a cluster is  primarily determined by the direction
of the last merger event. To  further clarify the relationship between
the ellipticity and the alignments note that smaller $n$ clusters will
tend to  grow by mergers of  small  mass clumps which  invariably fall
into   the  cluster along   a filament  (see    the video accompanying
Kauffmann \& Melott 1992). This will tend  to produce a high degree of
directionality in such models. For  large $n$ the infall is  basically
random but with larger mass clumps. This will produce a high degree of
randomness in the alignment angles for such models and more aspherical
clusters.

Cluster-cluster correlation, defined  as two clusters tending to point
in the   same direction, appears to also   increase as the   amount of
large-scale  power is increased in  the  initial conditions (i.e. with
decreasing  $n$). However, correlation    is present more  weakly than
alignment.

Our results  generally paint a  picture in  which background cosmology
has  less effect on  cluster morphology  than  has been hoped  by some
wishing to use  it as a probe.  As emphasized by White (1996), violent
relaxation  tends  to   move distributions  to   a somewhat  universal
distribution of shapes largely independent of background cosmology.

On  the positive side, both the  mean  ellipticity of clusters and the
scale  dependence of their axis  alignment seem  to reliably probe the
slope  of the primordial  power spectrum,  independent of $\Omega$. We
recommend focus    on these quantities  in  observational  studies and
analysis of simulations of candidate scenarios.

\section{Acknowledgments}

We wish  to thank  Jennifer Pauls   for her help.   This research  was
supported    by NASA grant    NAGW-2832  and Research  Experiences for
Undergraduate   support  under  NSF   grant   AST--9021414.  $N$--body
simulations were performed at the  National Center for  Supercomputing
Applications, Urbana,    Illinois.   ALM acknowledges  the    Barry M.
Goldwater Scholarship and both ALM and JT  acknowledge support of this
research from the University of Kansas's Undergraduate Research Award.
RJS thanks the Center for Computational  Sciences at the University of
Kentucky  for financial  support.

\newpage
\section{Figure Captions}

\noindent Figure 1  (a) The axial ratio  $e_2$ at two different stages
in the simulations,   for various spectral  indices  and low  or  high
$\Omega$ as  described in the text. (b)  The same as (1a) except axial
ratio $e_3$.  (c) The same  as (1a), except  the  axial ratio $e$  for
two--dimensional projected clusters.

\noindent Figure 2 (a) The axial ratio  $e_2$ plotted as a function of
radius  for    an  $n=-1$, $\Omega=1$  model   with  $b=1$  evolved to
$k_{nl}=4$, a $n=-1$ model with low $\Omega$  and $b=1$ evolved to the
same stage, and a $n=-1$ model evolved to  $k_{nl}=8$ with $b=2$.  (b)
The same  as  (2a) except axial  ratio  $e_3$. (c)  The same as  (2a),
except the axial ratio $e$ for two--dimensional projected clusters.

\noindent figure 3 A scatter plot of cos $\alpha$, the alignment angle
for cluster pairs.

\noindent Figure  4 A scatter   plot of cos $\beta$,  the  correlation
angle for cluster pairs.

\newpage

\pagestyle{empty}
\onecolumn

\centerline{TABLE 1}

\centerline{Boxes Sizes for Different Evolutionary Stages}

\begin{center}
\begin{tabular}{ccc} \hline \hline
$k_{nl}$ & Initial Spectral Index & Box Size (${\rm h^{-1}}$ Mpc)
\\
\cline{1-3}
8 & -2  &  300 \\
" & -1  &  270 \\
" &  0  &  240 \\
" & +1  &  220 \\
4 & -2  &  150 \\
" & -1  &  130 \\
" &  0  &  120 \\
" & +1  &  110 \\
\cline{1-3}
\end{tabular}
\end{center}

\clearpage

\setlength{\oddsidemargin}{0.0in}
\setlength{\evensidemargin}{0.0in}

\centerline{TABLE 2}

\centerline{Cluster Alignments in Three Dimensions}

\hspace{-0.75in}
\begin{tabular}{cccccc} \hline \hline 
{$0 \ Mpc/h < D < 15 Mpc/h$} & & $\Omega_0 < 1$	   &	& $\Omega_0 = 1$     \\
Spectral Index & $\Omega_0$ & $<cos\: \alpha>$        &  Significance Level & $<cos\: \alpha>$&   Significance Level\\ \hline 
-2 & 0.707 & 0.58 $\pm$ 0.03 & 0.003             & 0.55 $\pm$ 0.03 & 0.02             \\
-1 & 0.609 & 0.50 $\pm$ 0.04 & 0.01              & 0.53 $\pm$ 0.04 & 0.04             \\
0  & 0.487 & 0.61 $\pm$ 0.03 & 0.008             & 0.54 $\pm$ 0.03 & 0.15             \\
+1 & 0.347 & 0.71 $\pm$ 0.02 & 0.006             & 0.55 $\pm$ 0.04 & 0.21             \\ \hline 

{$15 \ Mpc/h < D < 30 Mpc/h$} & & $\Omega_0 < 1$   &	& $\Omega_0 = 1$     \\
Spectral Index & $\Omega_0$ & $<cos\: \alpha>$ & Significance Level & $<cos\: \alpha>$&   Significance Level\\ \hline 
-2 & 0.707 & 0.60 $\pm$ 0.04 & $10^{-7}$         & 0.59 $\pm$ 0.03 & $10^{-8}$            \\ 
-1 & 0.609 & 0.60 $\pm$ 0.04 & $10^{-6}$         & 0.57 $\pm$ 0.03 & 0.004            \\ 
0  & 0.487 & 0.55 $\pm$ 0.03 & 0.04              & 0.51 $\pm$ 0.03 & 0.83           \\
+1 & 0.347 & 0.52 $\pm$ 0.03 & 0.38              & 0.53 $\pm$ 0.03 & 0.32            \\ \hline 

{$30 \ Mpc/h < D < 60 Mpc/h$} & & $\Omega_0 < 1$    &  &$\Omega_0 = 1$     \\
Spectral Index & $\Omega_0$ & $<cos\: \alpha>$ & Significance Level & $<cos\: \alpha>$ &   Significance Level\\ \hline 
-2 & 0.707 & 0.55 $\pm$ 0.03 & $10^{-7}$         & 0.57 $\pm$ 0.03 & 10$^{-11}$ \\
-1 & 0.609 & 0.56 $\pm$ 0.03 & $10^{-5}$         & 0.52 $\pm$ 0.03 & 0.01            \\ 
0  & 0.487 & 0.50 $\pm$ 0.03 & 0.80              & 0.50 $\pm$ 0.02 & 0.81            \\
+1 & 0.347 & 0.50 $\pm$ 0.03 & 0.79              & 0.53 $\pm$ 0.03 & 0.21            \\ \hline 
\end{tabular}\\[0.5ex] 

\newpage

\centerline{TABLE 3}

\centerline{Cluster Correlations in Three Dimensions}

\hspace{-0.75in}
\begin{tabular}{cccccc} \hline \hline 
{$0 \ Mpc/h < D < 15 Mpc/h$} & & $\Omega_0 < 1$     &    &  $\Omega_0 = 1$     \\
Spectral Index & $\Omega_0$ & $<cos\: \beta>$& Significance Level & $<cos\: \beta>$ &  Significance Level\\ \hline
-2 & 0.707 & 0.55 $\pm$ 0.04 & 0.09      & 0.58 $\pm$ 0.03  & 0.02             \\
-1 & 0.609 & 0.44 $\pm$ 0.03 & 0.0009    & 0.48 $\pm$ 0.03  & 0.008             \\
0  & 0.487 & 0.50 $\pm$ 0.03 & 0.36      & 0.48 $\pm$ 0.03  & 0.64            \\
+1 & 0.347 & 0.50 $\pm$ 0.03 & 0.72      & 0.53 $\pm$ 0.04  & 0.16             \\ \hline

{$15 \ Mpc/h < D < 30 Mpc/h$} & & $\Omega_0 < 1$        &   &	$\Omega_0 = 1$     \\
Spectral Index & $\Omega_0$ & $<cos\: \beta>$ &  Significance Level & $<cos\: \beta>$ &   Significance Level\\ \hline 
-2 & 0.707 & 0.53 $\pm$ 0.03 & 0.01      & 0.54 $\pm$ 0.03  & 0.005  \\
-1 & 0.609 & 0.48 $\pm$ 0.03 & 0.003     & 0.58 $\pm$ 0.03  & 0.$10^{-5}$  \\
0  & 0.487 & 0.54 $\pm$ 0.03 & 0.01      & 0.47 $\pm$ 0.03  & 0.01  \\
+1 & 0.347 & 0.50 $\pm$ 0.04 & 0.10      & 0.50 $\pm$ 0.03  & 0.92  \\ \hline 

{$30 \ Mpc/h < D < 60 Mpc/h$} & & $\Omega_0 < 1$        &   &	$\Omega_0 = 1$     \\
Spectral Index & $\Omega_0$ & $<cos\: \beta>$ &  Significance Level & $<cos\: \beta>$ &  Significance Level\\ \hline 
-2 & 0.707 & 0.51 $\pm$ 0.03 & 0.018     & 0.54 $\pm$ 0.03  & $10^{-5}$             \\
-1 & 0.609 & 0.49 $\pm$ 0.03 & 0.003     & 0.51 $\pm$ 0.03  & 0.03            \\ 
0  & 0.487 & 0.49 $\pm$ 0.03 & 0.27      & 0.51 $\pm$ 0.02  & 0.11             \\
+1 & 0.347 & 0.50 $\pm$ 0.03 & 0.42      & 0.50 $\pm$ 0.03  & 0.37              \\ \hline
\end{tabular}\\[0.5ex] 

\newpage

\centerline{TABLE 4}

\centerline{Cluster Alignments in Two Dimensions}

\hspace{-0.75in}
\begin{tabular}{cccccc} \hline \hline 
{$0 \ Mpc/h < D < 15 Mpc/h$} & & $\Omega_0 < 1$ 	   &	& $\Omega_0 = 1$     \\
Spectral Index & $\Omega_0$ & $<\alpha>$        &  Significance Level & $<\alpha>$&   Significance Level\\ \hline 
-2 & 0.707 & 44.7$^\circ$ $\pm$ 1.5$^\circ$ & 0.08      & 43.5$^\circ$ $\pm$ 1.5$^\circ$ & 0.02 \\
-1 & 0.609 & 48.2$^\circ$ $\pm$ 1.4$^\circ$ & 0.002     & 45.3$^\circ$ $\pm$ 1.5$^\circ$ & 0.09 \\
0  & 0.487 & 43.7$^\circ$ $\pm$ 1.5$^\circ$ & 0.03      & 44.6$^\circ$ $\pm$ 1.5$^\circ$ & 0.82 \\
+1 & 0.347 & 40.0$^\circ$ $\pm$ 1.5$^\circ$ & $10^{-8}$ & 44.6$^\circ$ $\pm$ 1.5$^\circ$ & 0.85 \\ \hline 

{$15 \ Mpc/h < D < 30 Mpc/h$} & & $\Omega_0 < 1$        &	& $\Omega_0 = 1$     \\
Spectral Index & $\Omega_0$ & $<\alpha>$ & Significance Level & $<\alpha>$&   Significance Level\\ \hline 
-2 & 0.707 & 42.4$^\circ$ $\pm$ 1.4$^\circ$ & 0.00004   & 42.0$^\circ$ $\pm$ 1.5$^\circ$ & (10)$^{-5}$ \\
-1 & 0.609 & 39.6$^\circ$ $\pm$ 1.4$^\circ$ & $10^{-5}$ & 43.3$^\circ$ $\pm$ 1.4$^\circ$ & 0.001 \\ 
0  & 0.487 & 45.2$^\circ$ $\pm$ 1.5$^\circ$ & 0.006      & 44.5$^\circ$ $\pm$ 1.5$^\circ$ & 0.60 \\
+1 & 0.347 & 46.4$^\circ$ $\pm$ 1.5$^\circ$ & 0.43     & 45.7$^\circ$ $\pm$ 1.4$^\circ$ & 0.15 \\ \hline 

{$30 \ Mpc/h < D < 60 Mpc/h$} & & $\Omega_0 < 1$        &  &$\Omega_0 = 1$     \\
Spectral Index & $\Omega_0$ & $<\alpha>$ & Significance Level & $<\alpha>$ &   Significance Level\\ \hline 
-2 & 0.707 & 43.3$^\circ$ $\pm$ 1.5$^\circ$ & $10^{-5}$ & 43.6$^\circ$ $\pm$ 1.4$^\circ$ & 0.001 \\
-1 & 0.609 & 41.6$^\circ$ $\pm$ 1.5$^\circ$ & $10^{-4}$ & 44.7$^\circ$ $\pm$ 1.5$^\circ$ & 0.2  \\
0  & 0.487 & 44.2$^\circ$ $\pm$ 1.5$^\circ$ & 0.1       & 45.3$^\circ$ $\pm$ 1.5$^\circ$ & 0.21 \\
+1 & 0.347 & 45.1$^\circ$ $\pm$ 1.5$^\circ$ & 0.44      & 45.1$^\circ$ $\pm$ 1.5$^\circ$ & 0.97 \\ \hline 
\end{tabular}\\[0.5ex] 

\newpage

\centerline{TABLE 5}

\centerline{Cluster Correlations in Two Dimensions}

\hspace{-0.75in}
\begin{tabular}{cccccc} \hline \hline 
{$0 \ Mpc/h < D < 15 Mpc/h$} & & $\Omega_0 < 1$       &   &  $\Omega_0 = 1$     \\
Spectral Index & $\Omega_0$ & $<\beta>$& Significance Level & $<\beta>$ &  Significance Level\\ \hline
-2 & 0.707 & 44.5$^\circ$ $\pm$ 1.6$^\circ$ & 0.004      & 41.9$^\circ$ $\pm$ 1.6$^\circ$  & 0.001  \\
-1 & 0.609 & 45.8$^\circ$ $\pm$ 1.6$^\circ$ & $10^{-11}$ & 44.3$^\circ$ $\pm$ 1.5$^\circ$  & 0.03 \\
0  & 0.487 & 43.9$^\circ$ $\pm$ 1.4$^\circ$ & 0.02       & 44.4$^\circ$ $\pm$ 1.5$^\circ$  & 0.13  \\
+1 & 0.347 & 43.9$^\circ$ $\pm$ 1.5$^\circ$ & 0.04       & 43.3$^\circ$ $\pm$ 1.4$^\circ$  & 0.10  \\ \hline

{$15 \ Mpc/h < D < 30 Mpc/h$} & & $\Omega_0 < 1$      &   &	$\Omega_0 = 1$     \\
Spectral Index & $\Omega_0$ & $<\beta>$ &  Significance Level & $<\beta>$ &   Significance Level\\ \hline 
-2 & 0.707 & 43.2$^\circ$ $\pm$ 1.5$^\circ$ & 0.0009     & 43.4$^\circ$ $\pm$ 1.4$^\circ$  & 0.002  \\
-1 & 0.609 & 44.4$^\circ$ $\pm$ 1.6$^\circ$ & 10$^{-7}$  & 43.3$^\circ$ $\pm$ 1.5$^\circ$  & 0.008 \\
0  & 0.487 & 42.6$^\circ$ $\pm$ 1.4$^\circ$ & 10$^{-7}$  & 44.1$^\circ$ $\pm$ 1.5$^\circ$  & 0.14  \\
+1 & 0.347 & 42.2$^\circ$ $\pm$ 1.5$^\circ$ & 10$^{-6}$  & 45.1$^\circ$ $\pm$ 1.5$^\circ$  & 0.05  \\ \hline 

{$30 \ Mpc/h < D < 60 Mpc/h$} & & $\Omega_0 < 1$        &   &	$\Omega_0 = 1$     \\
Spectral Index & $\Omega_0$ & $<\beta>$ &  Significance Level & $<\beta>$ &  Significance Level\\ \hline 
-2 & 0.707 & 45.2$^\circ$ $\pm$ 1.6$^\circ$ & 10$^{-5}$  & 43.5$^\circ$ $\pm$ 1.5$^\circ$  & 0.0001  \\
-1 & 0.609 & 45.0$^\circ$ $\pm$ 1.5$^\circ$ & 10$^{-9}$  & 43.3$^\circ$ $\pm$ 1.5$^\circ$  & 0.0003  \\
0  & 0.487 & 45.7$^\circ$ $\pm$ 1.5$^\circ$ & 0.1     & 44.1$^\circ$ $\pm$ 1.5$^\circ$  & 0.0003   \\
+1 & 0.347 & 45.3$^\circ$ $\pm$ 1.4$^\circ$ & 0.32       & 45.2$^\circ$ $\pm$ 1.4$^\circ$  & 0.06   \\ \hline
\end{tabular}\\[0.5ex] 

\begin{figure}[t]
   \epsfxsize = 5.0truein
   \hskip 1.5truein
   \epsfbox{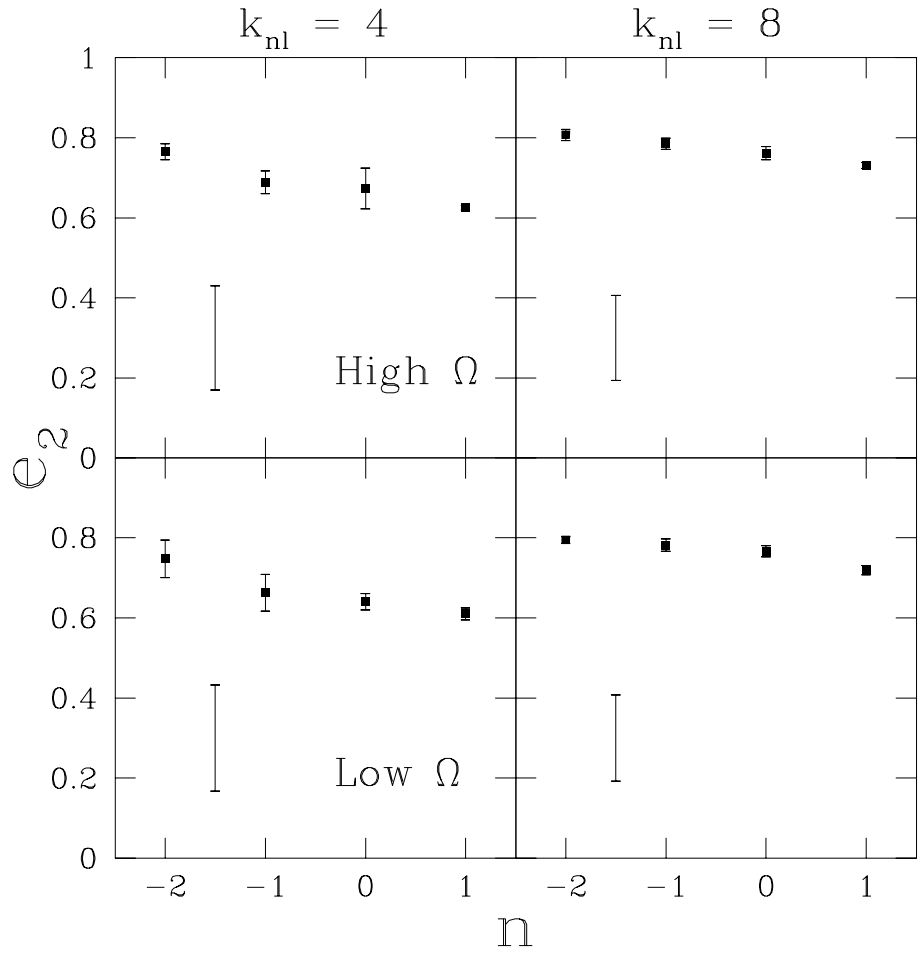}
   \caption{ Figure 1 (a) The  axial ratio $e_2$  at two different stages
             in the  simulations,  for various spectral  indicies and   
             low or high $\Omega$ as described  in the text.}
\end{figure}

\begin{figure}[t]
   \epsfxsize = 5.0truein
   \hskip 1.5truein
   \epsfbox{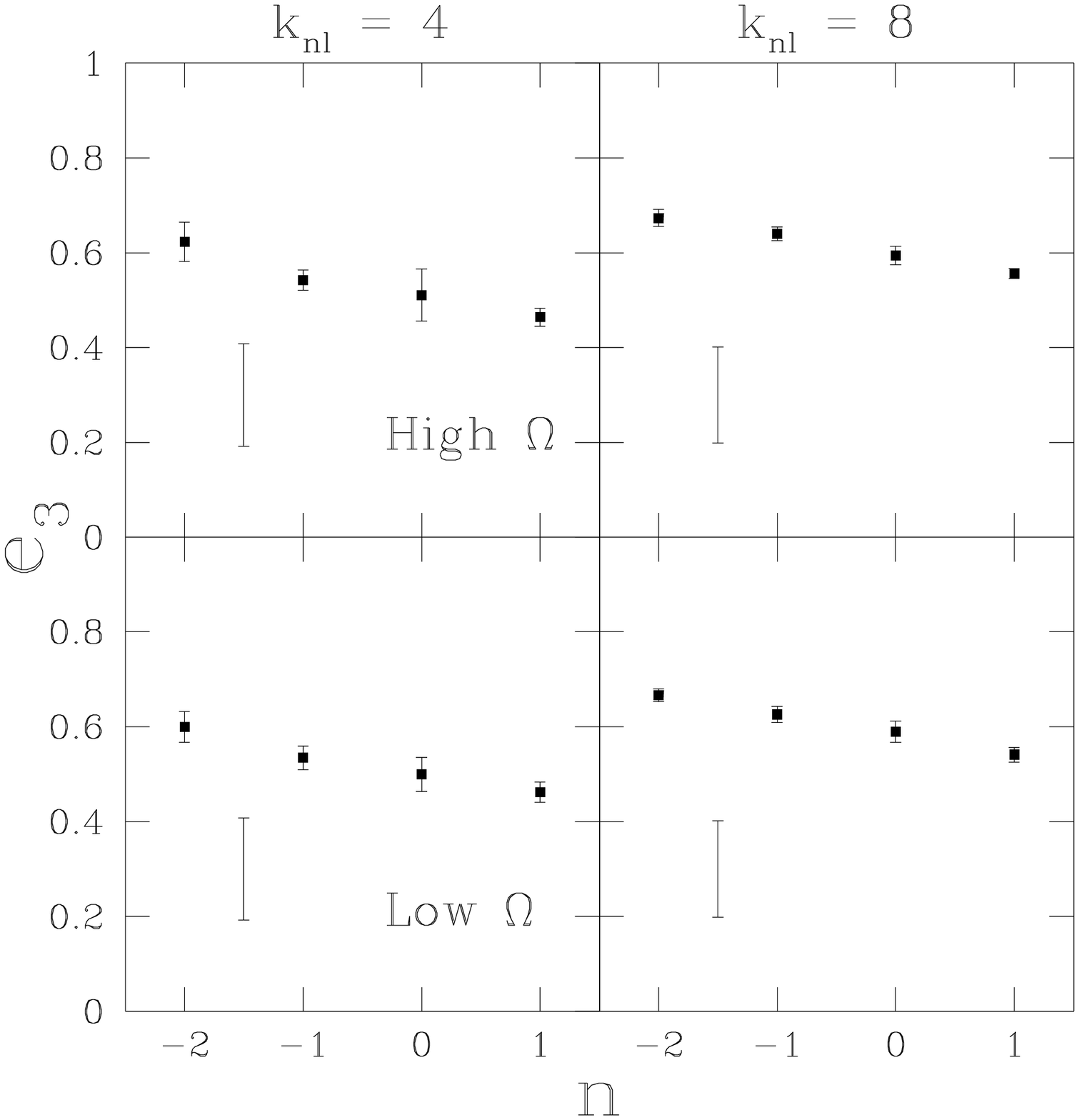}
   \caption{ Figure 1 (b)  The same as (1a) except axial
             ratio $e_3$.}
\end{figure}

\begin{figure}[t]
   \epsfxsize = 5.0truein
   \hskip 1.5truein
   \epsfbox{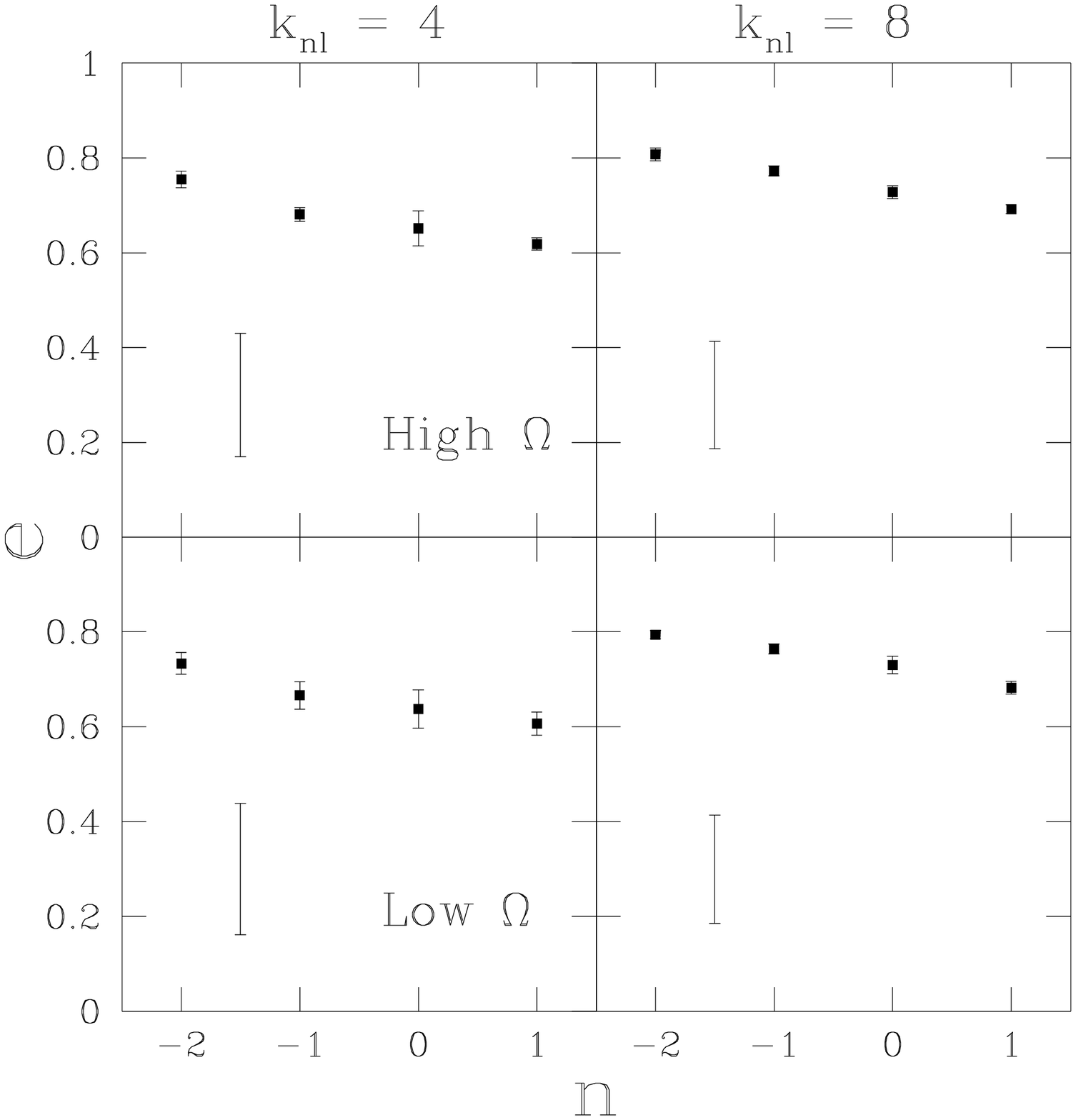}
   \caption{ Figure (c) The same  as (1a),  except the  axial ratio $e$  for
             two--dimensional projected clusters.}
\end{figure}

\begin{figure}[t]
   \epsfxsize = 5.0truein
   \hskip 1.5truein
   \epsfbox{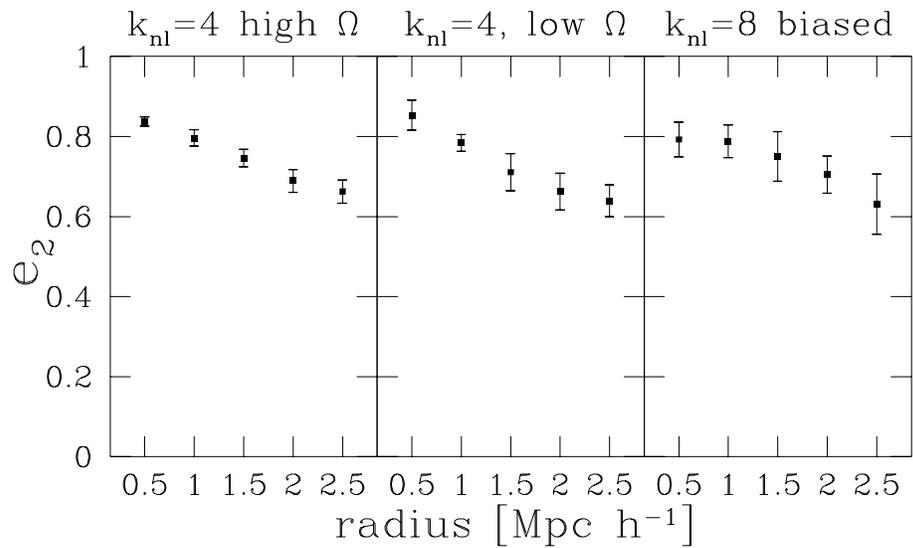}
   \caption{ Figure 2 (a) The axial ratio $e_2$  plotted as a function of
             radius   for  an $n=-1$,    $\Omega=1$   model with  $b=1$ 
             evolved  to $k_{nl}=4$, a $n=-1$ model with low $\Omega$  
             and $b=1$ evolved to the same stage, and a $n=-1$ model 
             evolved to  $k_{nl}=8$ with $b=2$.}
\end{figure}

\begin{figure}[t]
   \epsfxsize = 5.0truein
   \hskip 1.5truein
   \epsfbox{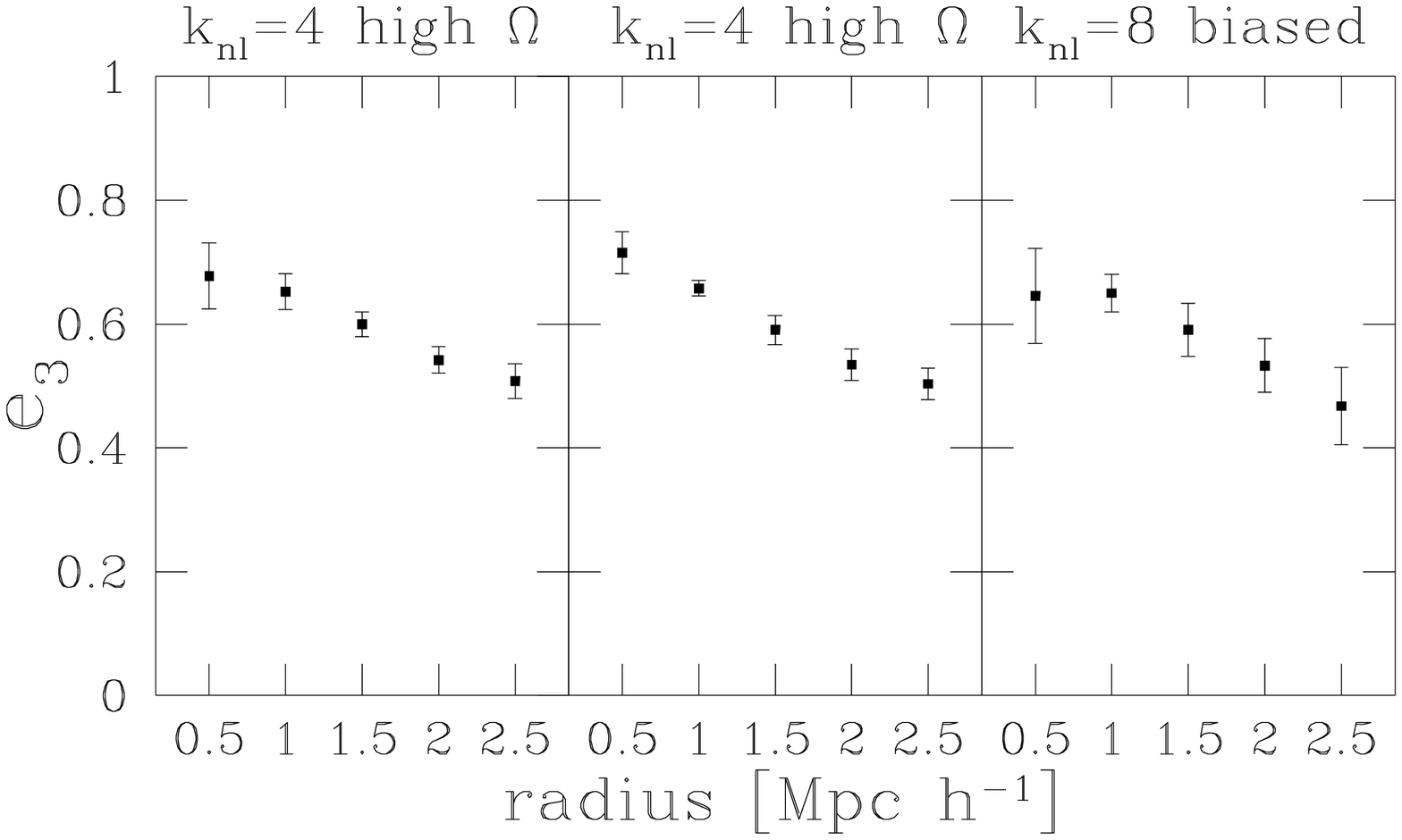}
   \caption{ Figure 2 (b) The same   as (2a) except  axial  ratio $e_3$. }
\end{figure}

\begin{figure}[t]
   \epsfxsize = 5.0truein
   \hskip 1.5truein
   \epsfbox{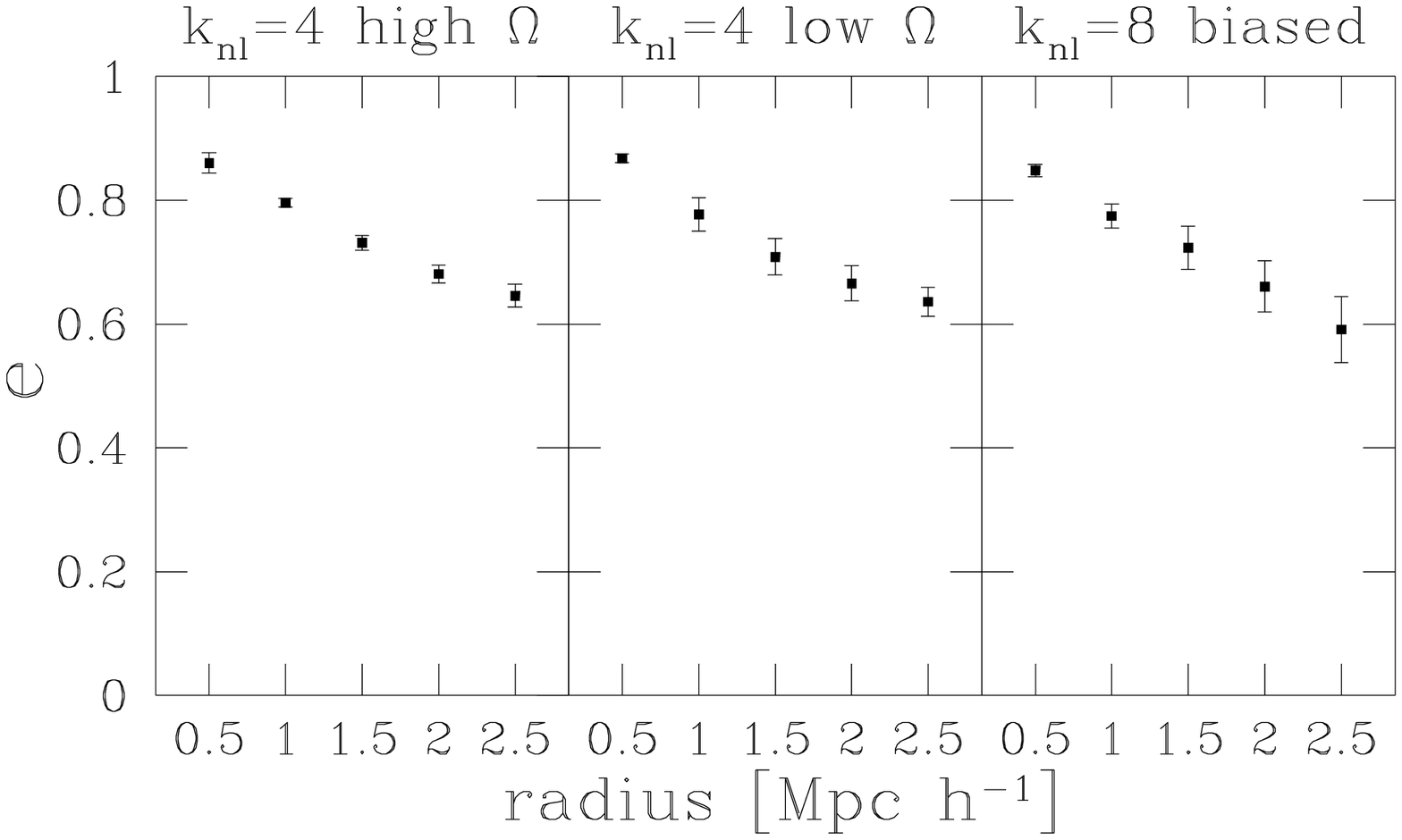}
   \caption{ Figure (c)  The same  as (2a), except the axial ratio $e$ for 
             two--dimensional projected clusters. }
\end{figure}

\begin{figure}[t]
   \epsfxsize = 5.0truein
   \hskip 1.5truein
   \epsfbox{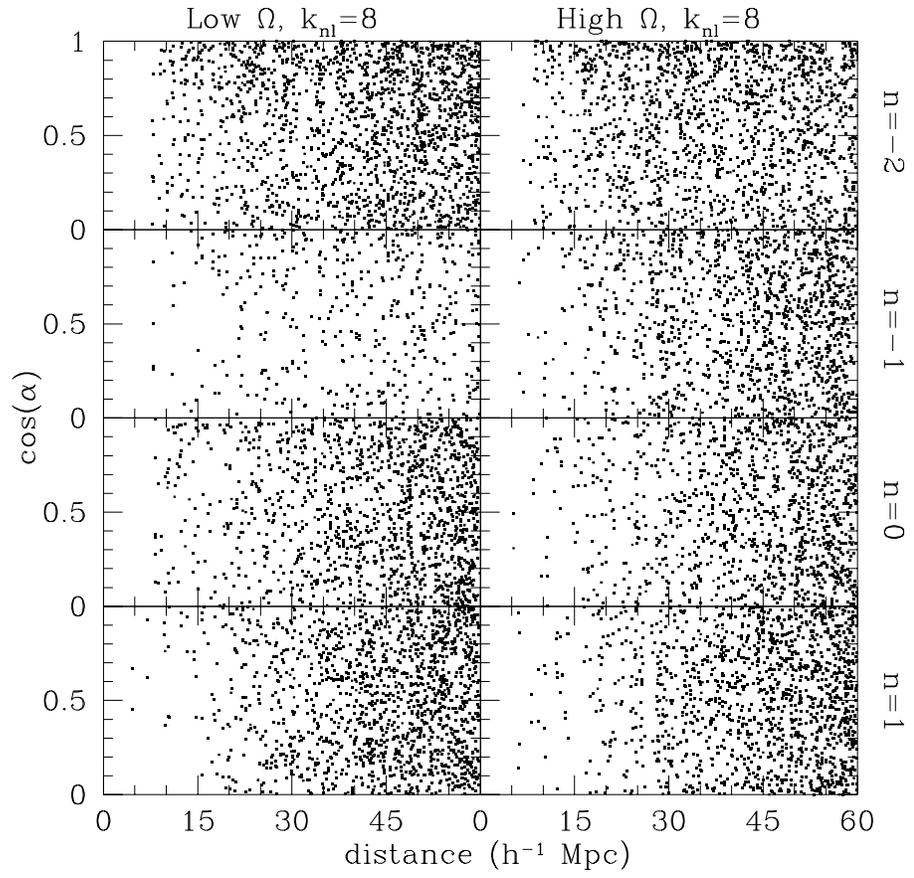}
   \caption{ Figure 3 A scatter plot of cos $\alpha$, the alignment angle
             for cluster pairs.}
\end{figure}
    
\begin{figure}[t]
   \epsfxsize = 5.0truein
   \hskip 1.5truein
   \epsfbox{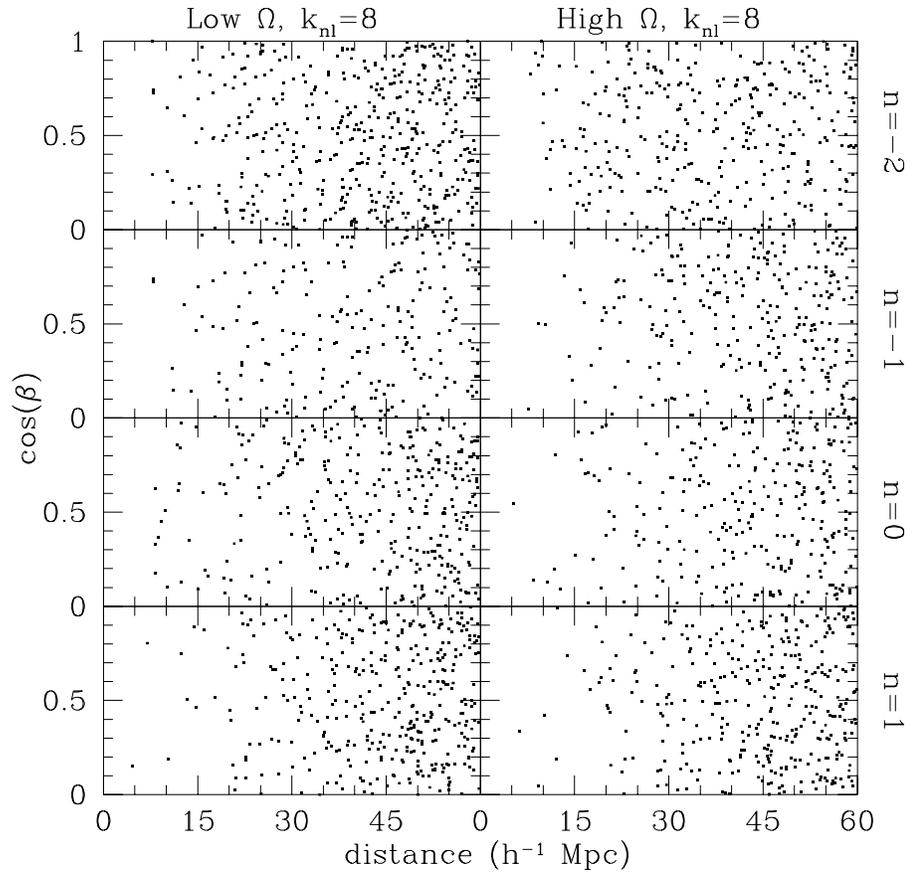}
   \caption{ Figure 4 A scatter   plot of  cos $\beta$, the  correlation
             angle for cluster pairs. }
\end{figure}

\end{document}